\documentstyle[epsf]{l-aa}
\def\be{\begin{equation}}
\def\ee{\end{equation}}
\def\Ec{E_{\rm C}}
\def\Eg{E_{\rm G}}
\def\Pc{P_{\rm C}}
\def\Pg{P_{\rm G}}
\def\Pgo{P_{{\rm G} 1}}
\def\Pgt{P_{{\rm G} 2}}
\def\Pco{P_{{\rm C} 1}}
\def\Pct{P_{{\rm C} 2}}
\def\gg{\gamma_{\rm G}}
\def\gc{\gamma_{\rm C}}
\def\ddt{{\partial\over\partial t}}
\def\Lg{\Lambda _{\rm G}}
\def\Lc{\Lambda _{\rm C}}
\def\Gg{\Gamma _{\rm G}}
\def\Gc{\Gamma _{\rm C}}
\begin{document}

\thesaurus{02(02.01.1;02.09.1;02.19.1)}


\title{Acoustic emission and corrugational instability of shocks
modified by strong particle acceleration}

\author{M. Mond\thanks{{\em Permanent address}: The Pearlstone Center for
Aeronautical Engineering Studies, Department of Mechanical Engineering,
Ben-Gurion University, Beer-Sheva, Israel} 
\and 
L. O'C. Drury}

\institute{Dublin Institute for Advanced Studies\\ 
5 Merrion Square\\
Dublin 2, Ireland} 

\date{Received;Accepted}

\maketitle

\markboth{M. Mond \& L. O'C. Drury}{Stability of shocks modified by
particle acceleration}

\begin{abstract}
The effect of particles that undergo strong
diffusive-shock-acceleration on the stability of the accelerating
shock is investigated. A two-fluid model is employed in which the
accelerated particles are treated as a fluid whose effect is
incorporated as an additional pressure in the momentum equation. The
Dyakov and Kontorovich stability criteria are used in order to study
the stability of those steady-state shocks that contain a gas
sub-shock. The downstream conditions of the latter are parametrized by
the ratio of the upstream acclerated-particles pressure to the total
pressure. For some range of values of that parameter, three possible
downstream states are possible for each upstream state. It is shown
that in that range of parameters the shocks are either corrugationally
unstable or lose their energy by spontaneous emission of acoustic as
well as entropy-vortex waves.
\end{abstract}

\section{Introduction}
Twenty years ago a process, now generally called diffusive shock
acceleration, was described whereby a significant part of the kinetic
energy flowing into an astrophysical shock could be used to accelerate
charged particles (Krymsky, \cite{Krymsky}; Bell 1978a,b; Axford 
et al., \cite{Axford}; Blandford \& Ostriker, \cite{Blan:Ost}). Clearly
the
reaction of
the accelerated particles on the shock dynamics and structure then
becomes an important effect. In the intervening years much work has
been done, but the problem is a difficult one and many aspects remain
open (see reviews by Jones \& Ellison, \cite{Jones}; Berezhko \& Krymsky,
\cite{Berezhko}; Blandford \& Eichler, \cite{Blan:Eich}; V\"olk,
\cite{Volk}). A
significant recent
advance is the approximate analytic theory of Malkov
(\cite{Malkova}).

For understandable reasons most of this work has assumed steady, or at
least quasi-stationary, shock structures and there has been relatively
little work on the question of shock stability. In this paper we apply
the classical criteria developed by Dyakov (\cite{Dyakov}) and Kontorovich
(\cite{Kontora}) for the corrugational instability of shock fronts and the
spontaneous emission of sound and entropy waves to the simplest model
of shocks strongly modified by particle acceleration, the two-fluid
model of Drury \& V\"olk (\cite{Druryb}). While this model has obvious
defects
(Heavens \& Drury, \cite{Heavens}; Achterberg et al., \cite{Achterberg})
and
has clear limitations (Drury et al., \cite{Druryc}) it has proven
to be a useful tool in interpreting numerical studies (Falle \&
Giddings, \cite{Falle}; Jones \& Kang, \cite{Jones:Kang}) when used with
appropriate
caution.

\section{The two-fluid model}

The key assumption in the two-fluid model is that the spatial
transport of the accelerated particles can be represented by diffusion
with a single effective diffusion coefficient. This enables one to
integrate over the particle energy spectra and regard the accelerated
particles as a second fluid with significant energy density and
pressure, but negligible inertia. For historical reasons the
accelerated particles are often identified as cosmic rays and the
energy density and pressure denoted $\Ec$ and $\Pc$,
\begin{eqnarray}
\Ec&=&\int 4\pi p^2 T(p) f(p) dp\\
\Pc&=&\int 4\pi p^2 {p v(p)\over 3} f(p) dp
\end{eqnarray}
where $p$ is the magnitude of the accelerated particle momentum,
$f(p)$ is the isotropic part of the phase space density, $T(p)$ the
associated kinetic energy and $v(p)$ the particle speed. 

The two fluid
model then consists of the standard hydrodynamical equations for the
background gas with density $\rho$, velocity $U$, energy density
$\Eg$ and pressure $\Pg$ 
including both pressures in the momentum equation
and a diffusive energy flux term with coefficient $\kappa$ 
in the internal energy equation for
the accelerated particle fluid
\begin{eqnarray}
\ddt\rho + \nabla\cdot\left(\rho {\bf U}\right) &=& 0,\\
\ddt {\bf U} + {\bf U}\cdot\nabla {\bf U}
+{1\over\rho}\nabla\left(\Pc+\Pg\right) &=&
0,\\
\ddt \Pg + {\bf U}\cdot\nabla\Pg + \gg\Pg\nabla\cdot {\bf U} &=& 0,\\
\ddt \Pc + {\bf U}\cdot\nabla\Pc + \gc\Pc\nabla\cdot {\bf U} &=&
\nabla\left(\kappa\nabla\Pc\right).  
\end{eqnarray}
At discontinuities this system has to be replaced by the usual
conditions of mass, momentum and energy conservation and continuity of
$\Pc$. The adiabatic exponents $\gg$ and $\gc$ are defined by
\be
P_{\rm G,C} = (\gamma_{\rm G,C}-1)E_{\rm G,C}.
\ee
For an ideal monatomic gas $\gg=5/3$ and $4/3<\gc<5/3$. 

The simplicity of the two-fluid model is rather deceptive. All
information about the energy (or momentum) spectrum of the accelerated
particles has been put in the two closure parameters $\kappa$ and
$\gc$ and, perhaps not surprisingly, the solutions turn out to depend
very sensitively on the assumptions made for these parameters. To some
extent this can be overcome by using simple physical models for
the spectral evolution to improve the estimates (Duffy et al.,
\cite{Duffy}). In addition Malkov \& V\"olk (\cite{Malkovb}) have shown
that
even in circumstances where the simple two-fluid model is invalid a
form of renormalised two-fluid model is applicable. 

Our approach in this paper is rather different, We regard the
two-fluid equations not as an approximation to some more complicated
system, but as the simplest model system (in currently popular
terminology a ``toy'' model) in which we can study the interaction
between particle acceleration and shock structure. We note that if the
accelerated particles were to have an energy independent diffusion
coefficient the two-fluid equations would be exact. Thus the model is
physical and does not violate any fundamental physical principles.

All possible stationary shock structures within this model were
classified by Drury \& V\"olk, (\cite{Druryb}). It is interesting to
recall
one of the concluding remarks in this paper. ``Perhaps the most
interesting, certainly the most characteristically nonlinear, feature
of this model is that under certain circumstances it has three
solutions. Whether all are stable and can occur as time asymptotic
states in a physically reasonable evolution remains to be
studied...''. This paper is a belated answer to this question.

\section{Stability criteria for shocks}

The study of the stability of shocks was first undertaken by Dyakov, 
(\cite{Dyakov}). He considered a planar shock wave propagating in an
unbounded
medium and investigated the evolution of small sinusoidal corrugations of the
shock front. Dyakov showed that, under certain conditions, 
perturbations that are confined to the front's area grow exponentially with
time. This phenomena is termed {\it corrugational instability}. 
Shocks that are corrugationally unstable can only be
short-lived because a reorganization of the entire flow will take place on a
time-scale of the order of the inverse growth rate.

In addition to the corrugational instability, Dyakov (\cite{Dyakov}) and
later
Kontorovich (\cite{Kontora},\cite{Kontorb}) determined the conditions
under
which small
acoustic as well as entropy-vortex perturbations in the form of
sinusoidal two dimensional waves can be emitted from the shock front.
This phenomenon is termed {\it spontaneous emission}. Even though the
spontaneously emitted waves do not grow with time, their occurence
may eventually render the flow behind the spontaneously emitting shock
unstable; this is because the shock's energy is continuously carried
away by the outgoing waves and this finally results in the reorganization
of the flow. 

Both corrugational instability and spontaneous emission may be treated as
resonant reflection of acoustic waves from the shock front where
the reflection coefficient of the acoustic waves from the shock is
infinite. Wereas historically the reflection coefficient has been
calculated in terms of the geometrical parameters of the problem,
such as the angles of incidence and reflection, Mond
\& Rutkevich (\cite{Monda}) have recently recast the traditional
Dyakov-Kontorovich formulation into a
frequency-dependent representation of the reflection coefficient. Such
a representation is convenient in solving various boundary value
problems such as, for example, investigation of the eigenperturbations
between the shock and some other reflecting surface behind it.
A further advantage of the frequency-dependent formulation
is evident when dissipative effects are taken into account
and the relevant frequencies become complex. In this case simple
geometric
concepts like the angle of incidence are of restricted physical meaning. 

The reflection coefficient is calculated by considering an incident
acoustic wave as well as reflected acoustic and entropy-vortex waves in
the region downstream from the shock of the form
\be
\delta f\exp(-i\omega t+i{\bf k}\cdot {\bf r}),
\label{eq:pertur}
\ee
where $\delta f$ is the amplitude of the perturbation of the
relevant physical quantity. While
$\omega $ and the component of ${\bf k}$ which is parallel to the
unperturbed shock front are the same for all the waves involved in the
process, the perpendicular component of ${\bf k}$ is determined
seperately
for each wave according to the appropriate dispersion relation. The
upstream perturbations are zero due to
the supersonic velocity there. The reflection coefficient is obtained 
by imposing the continuity of
the tangential velocity component as well as the appropriate jump
condition of the normal velocity components at the perturbed shock
front. Its frequency-dependent representation is given by: 
\be
{\cal R} =-\frac{f(q)-(q^2-1)^{1/2}}{f(q)+(q^2-1)^{1/2}}
\label{eq:Refcoeff}
\ee
where
\be
f(q)=\frac{1-h}{2M_2}q-\frac{(1+h)\eta
M_2}{2q(1-M_2^2)},\;\;\; q=\frac{\omega }{k_{\Vert}c_2 (1-M_2^2)^{1/2}},
\label{eq:qdef}
\ee
$M_2$ is the downstream Mach number and $\omega$ and $k_{\Vert}$ are
the frequency, and the wave vector component parallel to the shock's
front, of the acoustic as well as of the entropy-vortex waves. As will
be seen later, the parameter $h$ plays a crucial role in determining
the shock's stability and is given by
\be
h=-v_2^2{\biggr(}\frac{\partial \rho_2 }{\partial p_2 } {\biggr )}_{\rho
_1 , p_1},
\label{eq:h}
\ee
where $v$, $\rho$ and $p$ are the velocity, density and pressure of
the gas, respectively and the subscripts 1 (2) denotes the
corresponding values in the upstream (downstream) side of the
shock. Thus, $h$ is given in terms of the derivative of the downstream
density with respect to the downstream pressure for fixed upstream
conditions, i.e., in terms of the slope of the Hugoniot curve.

The problem of stability may be cast now into the following question:
under what conditions does the denominator on the right hand side of
eq.~(\ref{eq:Refcoeff}) become zero? The answer is that two
families of two dimensional acoustic as well as entropy-vortex waves
may give rise to an infinite reflection coefficient under two distinct
conditions. The first family of waves is characterized by purely
imaginary $\omega$ and purely imaginary $k_{\bot}$ (the wave vector
component perpendicular to the shock's front). Hence, such waves grow
exponentially in time and decay exponentially away from the
shock. They give rise to the corrugational instability and the
conditions for their existence are
\be
h<-1, \qquad\hbox{or}\qquad h>1+2M_2.
\label{eq:Dyakov}
\ee
When condition (\ref{eq:Dyakov}) is satisfied, the requirement of an
infinite reflection coefficient results in a linear dispersion
relation between the growth rate of the instability and the
corrugation wavelength, $k_{\Vert}$.

The second family of waves that can give rise to an infinite
reflection coefficient is characterized by real frequencies and real
wave vectors.  Those waves are the manifestation of spontaneous
emission and the condition for their occurence is
\be
h_c<h<1+2M_2
\label{eq:SE}
\ee
where
\be
h_c=\frac{1-(1+\eta )M_2^2}{1+(\eta -1)M_2^2}
\label{eq:hc}
\ee
where $\eta =\rho _2 /\rho _1$ is the compression ratio.  The
implication of condition (\ref{eq:SE}) is that if it is satisfied,
there is a one parameter family containing an infinite number of two
dimensional acoustic as well as entropy-vortex waves any of which can
be spontaneously emitted from the shock front. A linear dispersion
relation exists between the frequencies of the spontaneously emitted
waves and their parallel wave vectors.

\section{Stability of shocks modified by particle acceleration}

When conditions (\ref{eq:Dyakov}) and (\ref{eq:SE}) are applied to
shocks in an ideal polytropic gas the result is absolute stability
against both corrugational instability and spontaneous wave
emission. This result is not unexpected and has been well known
experimentally since the early days of shock wave research. However,
as nonideal processes become progressively more important, the
stability properties of the flow may significantly change and the
shock may become susceptible to one of the instabilities discussed in
the previous section. An example for such occurence has been recently
discussed in Mond \& Rutkevich (\cite{Mondb}) where it has been shown that
strong ionizing shocks spontaneously emit acoustic waves if their Mach
number exceeds a certain value.

Here, the effect of shock accelerated particles on the stability of
the accelerating shock is investigated. For that purpose, the
two-fluid model discussed in section 2 is employed. Within the
framework of this model, a shock is a transition layer between two
uniform states, whose length is determined by the mean diffusion
coefficient of the particles. Following Drury (1983), it can be shown
that the transition layer may either be smooth or contain a sub-shock
in the background gas. It is the latter case that will be the focus of
the current investigation.

For the purpose of this stability analysis the transition layer that
includes the gas sub-shock will be regarded as a single surface of
discontinuity. It is this surface that will be called the shock from
now on. Such a procedure is admissible if the wavelengths of the
investigated perturbations are much larger than the thickness of the
transition layer.

\subsection{The Hugoniot curve}

It is obvious from equations (\ref{eq:h})-(\ref{eq:SE}) that the shape
of the Hugoniot curve plays a crucial role in determining the
stability properties of shocks that propagate into a given uniform
state.  In order to obtain the Hugoniot curve, the conservation
equations that relate the flow variables on both uniform-state-sides
of the shock are written in a frame of reference that is moving with
the shock's velocity. For that purpose it is convenient to introduce
the following nondimensional variables:
\be
\eta=\frac{\rho _2}{\rho _1},\;\;\;  
\xi=\frac{\Pco}{\Pgo +\Pco},
\label{eq:nondim1}
\ee
and
\be
\Lg=\frac{\Pgt}{\Pgo +\Pco},
\;\;\;  \Lc=\frac{\Pct}{\Pgo +\Pco},
\label{eq:nondim2}
\ee  
where the subscript 1 (2) denotes the values of the upstream (downstream)
corresponding variables. Setting all time derivatives to zero in equations
(3)-(6), the following equations are obtained:
\begin{eqnarray}
\label{eq:moms}
J^2(1-\frac{1}{\eta})-\Lg-\Lc &=&-1,\\
\label{eq:eners}
\frac{1}{2}J^2(1-\frac{1}{\eta ^2})-\Gg \frac{\Lg}{\eta}-\Gc
\frac{\Lc}{\eta} &=& -\theta,\\
\label{eq:momg}
J^2(\frac{1}{\eta _a}-\frac{1}{\eta })+\Lambda _a -\Lg &=&0,\\
\label{eq:energ}
\frac{1}{2}J^2(\frac{1}{\eta _a ^2}-\frac{1}{\eta ^2})+\Gg (\frac{\Lambda
_a}{\eta _a}-\frac{\Lg}{\eta}) &=&0,\\
\label{eq:adiabat}
\Lambda _a &=&\eta _ a ^{\gg}.
\end{eqnarray}
where $\theta = (\Gg (1-\xi)+\Gc \xi)$, $\Gamma _{G,C}=\gamma
_{G,C}/(\gamma _{G,C}-1)$ and $J$ is the
normalized mass flux and is given by
\be
J^2=\frac{\rho _1 U _1 ^2 }{\Pgo+\Pco}.
\label{eq:flux}
\ee
Equations (\ref{eq:moms}) and (\ref{eq:eners}) represent the momentum and
energy conservation, respectively, across the shock while equations
(\ref{eq:momg}) and (\ref{eq:energ}) represent the conservation of the
background gas momentum and energy across the gas sub-shock.
The variables $\Lambda _a$ and $\eta _a$ are the normalized pressure and
density at the foot of the gas sub-shock and are adiabaticaly related to
the upstream conditions according to eq. (\ref{eq:adiabat}).

The solution of the system of equations (\ref{eq:moms})-(\ref{eq:adiabat})
result in a one parameter ($\xi$) family of Hugoniot curves. Two such
typical curves for $\gg =5/3$ and $\gc =4/3$ are shown in Fig. 1 for
$\xi=0.45$ and in Fig. 2 for $\xi=0.1$. As can
be
seen in both figures, for small values of $M_1$ the Hugoniot curve follows
its
single-fluid, $\gamma =5/3$ counterpart while for large values of $M_1$ it
is cosmic-rays dominated as it asymptotically approaches the value of $7$
as predicted for a single-fluid with $\gamma =4/3$. It should be noted,
however, that for high enough values of $M_1$ no physically acceptable
solution of the system of equations (\ref{eq:moms})-(\ref{eq:adiabat})
exists which indicates that only a smooth transition layer (without a gas
sub-shock) may exist for that range of parameters. The two asymptotic
parts of the Hugoniot curve are connected by an intermediate section. As 
can be seen in Fig. 2, for small values of $\xi$ for some portion of the 
intermediated section there are three possible downstream states for
each upstream state. As
will be seen later on, the shock instabilities occur at that intermediate
range of parameters. 

\begin{figure}
\epsfxsize=\hsize
\epsfbox{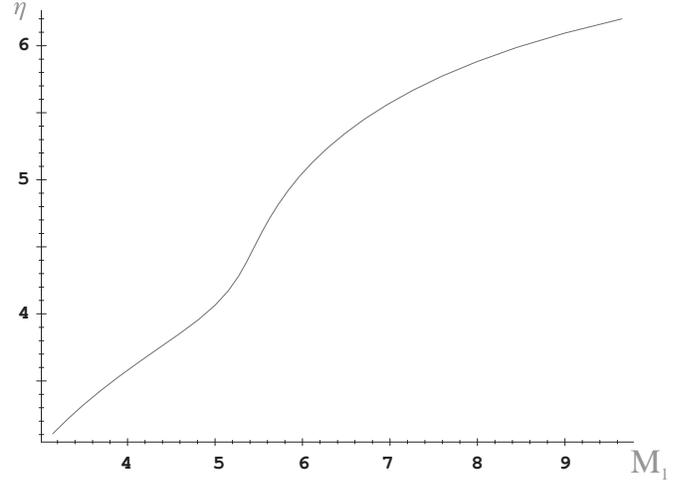}
\caption[]{The Hugoniot curve for $\xi =0.45$}
\end{figure}

\begin{figure}
\epsfxsize=\hsize
\epsfbox{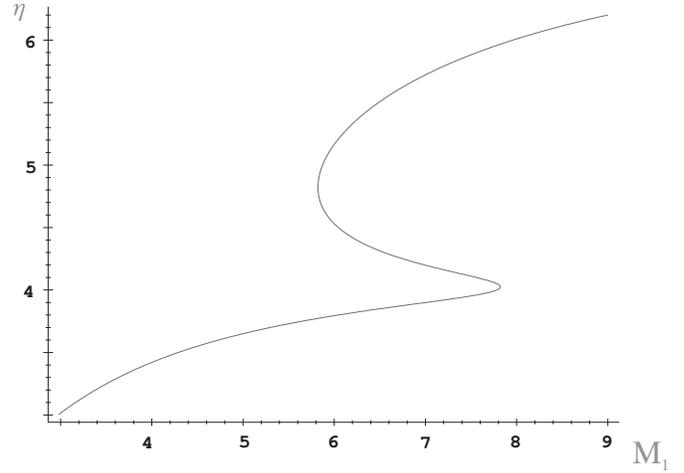}
\caption[]{The Hugoniot curve for $\xi =0.1$}
\end{figure}

\subsection{The stability analysis}
As was discussed in the previous section the concept of a single surface
of discontinuity is meaningful only for waves whose wavelengths are much
longer than the thickness of the transition layer. In that long
wavelength limit the
acoustic perturbations in the background gas couple to the accelerated
particles and travel at the enhanced speed
\be
c^2=\frac{\gg \Pg +\gc \Pc}{\rho}
\label{eq:soundspeed}
\ee
In this case the parameters $h$ and $M_2$ that are needed for the
calculation of the
stability criteria (\ref{eq:Dyakov}) and (\ref{eq:SE}) may be expressed in
terms of the nondimensional variables defined in Eqs. (\ref{eq:nondim1})
and (\ref{eq:nondim2})  as
\be
h=-\frac{J^2}{\eta} \frac{\partial \eta}{\partial \Lambda}
\label{eq:nondimh}
\ee
where $\Lambda =\Lg +\Lc$ is the normalized total downstream pressure,
and
\be
M_2^2=\frac{J^2}{\eta (\Gg \Lg +\Gc \Lc)}.
\label{eq:Mach}
\ee

The variables $h$ and $M_2$ were calculated according to
(\ref{eq:nondimh}) and(\ref{eq:Mach}) along the Hugoniot curves and then 
were used in criteria (\ref{eq:Dyakov}) and (\ref{eq:SE}). The
results are
shown in Fig.~3 and ~4 where $h_c -h$ is plotted for $\xi=0.45$ and
$\xi=0.1$, respectively. 

\begin{figure}
\epsfxsize=\hsize
\epsfbox{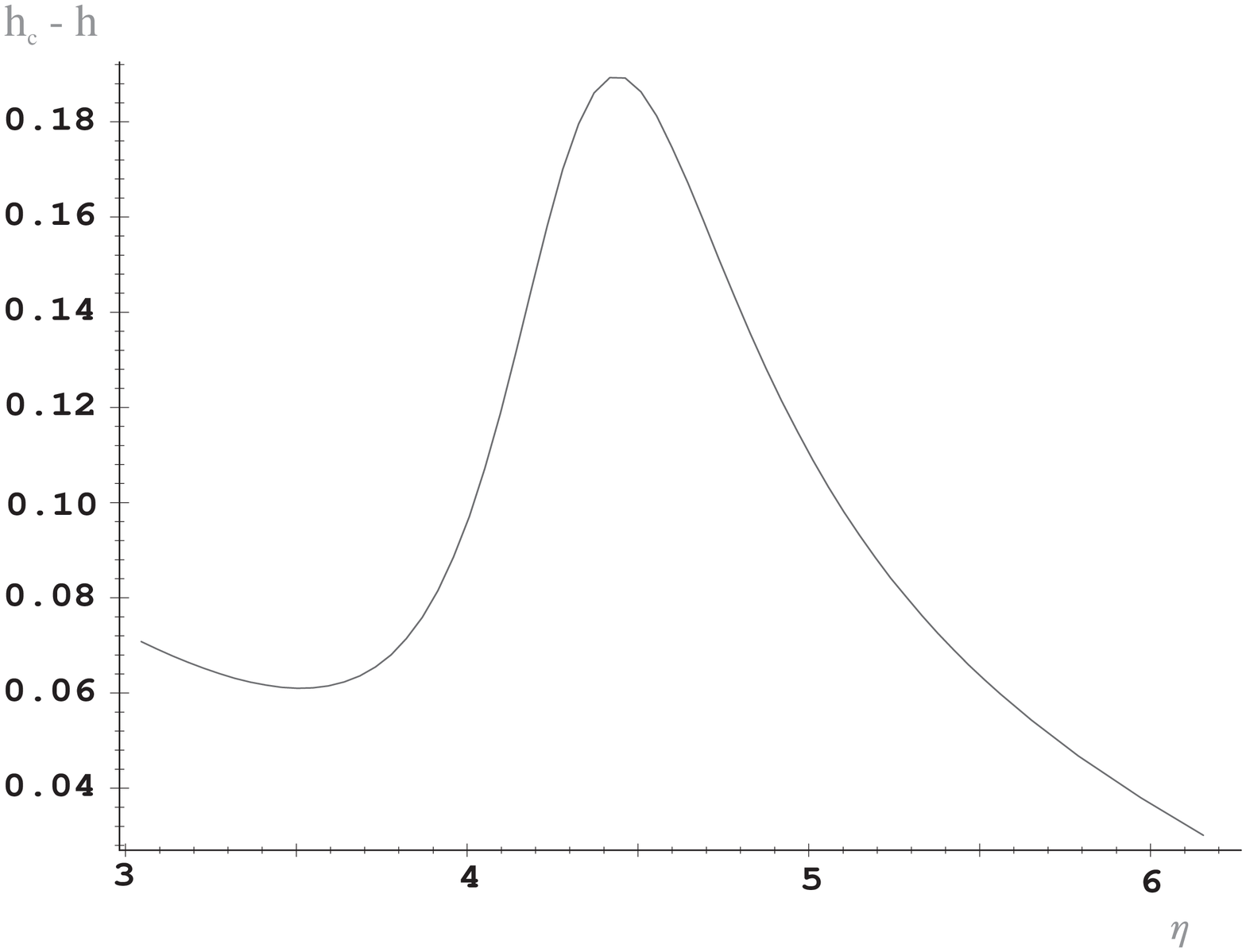}
\caption[]{The behaviour of $h_c -h$ as a function of $\eta$ for $\xi
=0.45$}
\end{figure}

\begin{figure}
\epsfxsize=\hsize
\epsfbox{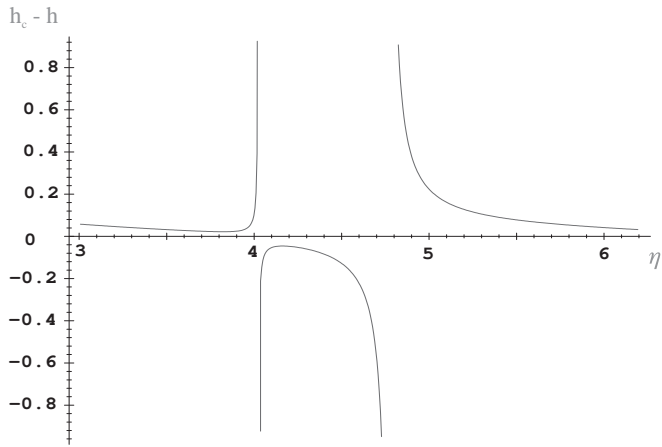}
\caption[]{The behaviour of $h_c -h$ as a function of $\eta$ for  $\xi
=0.1$}
\end{figure}

It is obvious that all possible
shocks for the $\xi =0.45$ case are stable against both corrugation
instability as well as spontaneous emission. This changes however for
$\xi =0.1$. In that case there are three possible downstream states for
each
upstream state at some intermediate section of the Hugoniot curve.
According to Fig.~4 shocks that belong to the descending part of the
intermediate section are unstable under spontaneous acoustic emission.
This is so since along that portion of the Hugoniot curve $h$ is positive
(see Eq. (\ref{eq:nondimh})) while $h_c$ remains negative.  Furthermore,
Fig.~2 indicates that the
shocks near the ``knees''
of the Hugoniot curve are corrugationally unstable. 
An alternative algebraic form for the parameter $h$ is easily seen to
be 
\be 
h = -1/\left(1 + 2(\eta-1){\eta\over U_1}{dU_1\over d\eta}\right).
\ee 
This shows that that $h= -1$, indicating the onset of
corrugational instability, at precisely those points where the graph
of $\eta$ against $U_1$ is vertical. A slight further inflection of the graph
then allows the denominator to become zero and $h\to\infty$. For
such high values of $h$ the dispersion relation that is obtained for the
corrugational instability is approximately given by
\be
q^2=-\frac{\eta M_2^2}{1-M_2^2},
\label{eq:dispers}
\ee
so that the growth rate according to the definition in eq. (\ref{eq:qdef})
is
\be
\sigma =\sqrt{\eta }M_2 k_{\Vert}c_2.
\label{eq:growth}
\ee

To complete the stability analysis it should be noted that in the short
wavelength limit the acoustic waves
decouple from the particles and propagate within the background gas at the
gas sound speed $(\gg \Pg /\rho )^{1/2}$. In this limit the acoustic waves
propagate through the transition layer and are reflected off the gas
sub-shock. Hence, the flow is stable under short wavelength perturbations.

\section{The effect of particle diffusion}
As was discussed in section 2 the spontaneously emitted waves are
marginally stable in the sense that their frequencies are purely real. It
is of interest, therefore, to investigate the effects of dissipative
mechanisms on the stability of the flow as they
give rise to imaginary parts of both the frequencies and the perpendicular
wave vector. Hence the question to be asked is whether, under the effects
of dissipation, some of the
spontaneously emitted eigenperturbations acquire positive imaginary
parts of both their eigenfrequencies and their perpendicular wave vectors.
This question is still an open one and no general satisfactory answers are
available. Furthermore, conflicting claims may be found in the literature
with respect to the stabilizing/destabilizing effects of dissipation.

In order to study the effect of the particles dissipation let us
reconsider the denominator of the right hand side of
Eq.~({\ref{eq:Refcoeff}). It was tacitly assumed before
that
the coefficients in that expression (and in particular $M_2$) were real.
Then the resulting eigenfrequencies were either purely imaginary
(corrugation instability) or purely real (spontaneous emission). However,
when the effects of particle diffusion are included this ceases
to be true and the Mach number, and consequently the frequencies of
the spontaneously emitted waves, become complex. Linearizing
Eqs. (3)-(7), assuming that the diffusion is small and that 
$k_{\Vert}>>k_{\bot}$ results in the
following expression for the sound speed
\be
c^2=c_0^2(1-i\frac{\gc \Pc }{\gg \Pg + \gc \Pc}x)
\label{eq:modsound}
\ee
where $c_0$ is the sound speed in the absence of diffusion and is given by
(\ref{eq:soundspeed}) and $x$ is defined by
\be
x=\frac{k_{\Vert}\kappa}{c_{01}}.
\label{eq:xdef}
\ee
With the above modified expression of the sound speed the Mach number is
given by
\be
M^2=M_0^2(1+i\frac{\gc \Pc }{\gg \Pg + \gc \Pc}x).
\label{eq:modMach}
\ee
Using expression (\ref{eq:modMach}) the denominator of the right hand side
of Eq. (\ref{eq:Refcoeff}) is equated now to zero and solutions for the
frequencies are searched
such that the corresponding imaginary parts of the perpendicular wave
vectors are positive. The sign of the imaginary part of the
perpendicular wave vector is checked by using the dispersion relation for
the acoustic waves that propagate away from the shock
\be
k_{\bot}=\frac{\omega [-M +  (1-q^{-2})^{1/2}]}{c(1-M^2)}.
\label{eq:kperp}
\ee
Using the results for a shock ($M_1 =4.5$) that belongs to the spontaneous
emission regim it is found that the particles dissipation give rise to the
damping of the spontaneously emitted waves. As is expected, the damping
rate is proportional to $k_{\Vert}^2$ and is much smaller that the
corresponding real part of the frequency. Thus, the energy of the shock is
continuously being carried away from the shock by the spontaneously
emitted waves and is subsequently deposited over a length scale of
$1/k_{\bot}$.

\section{Discussion}

This paper fills a significant gap in the theoretical understanding of
the two fluid model by demonstrating what had previously only been
conjectured, that in those cases where the model produces multiple
solutions the intermediate solution is unstable. It is also, as far as
we know, the only case where Dyakov's criterion for corrugational
instability is actually realised in a physically plausible system.  In
addition to its theoretical interest, we note that the two fluid model
is quite extensively used as the basis for numerical studies of
particle acceleration effects in astrophysical systems; clearly the
knowledge that shock instabilities of the type described here can,
indeed must, occur in such simulations is an important piece of
information. It will be interesting to see whether similar
instabilties are found in more realistic models with momentum
dependent diffusion.  As we have already noted the two fluid model
captures much of the correct physics and is not unphysical. This,
together with Malkov's renormalisation interpretation of the two fluid
model, suggests that these effects should also occur in the more
complicated models.

\begin{acknowledgements}

This work was carried out while MM was in receipt of a Senior Marie
Curie fellowship (contract ERBFMBICT971894) from the European Union
under the programme for Training and Mobility of Researchers; MM 
thanks the school of cosmic physics of the Dublin Institute for Advanced
Studies for its friendly and warm hospitality. 

\end{acknowledgements}

\end{document}